\documentclass[twocolumn,amsmath,amssymb,superscriptaddress]{revtex4}

\usepackage{graphicx}
\usepackage{bm}
\usepackage{times}

\begin{document}

\title{Sobre las Relaciones de Incerteza de Heisenberg entre Tiempo y Energ\'{\i}a: Una nota did\'actica}

\author{Gast\'on Giribet}

\medskip

\affiliation{Departamento de F\'{\i}sica, Facultad de Ciencias Exactas y Naturales de la Universidad de Buenos Aires. 
             Ciudad Universitaria, Pab. I, c.p. 1428, Buenos Aires, 
             Argentina}

\affiliation{Instituto de F\'{\i}sica de La Plata, Universidad Nacional de La Plata. 
             La Plata, Argentina}

\begin{abstract}
Esta nota fue escrita originalmente para los alumnos del curso introductorio a la mec\'anica cu\'antica dictado en la Facultad de Ciencias Exactas y Naturales de la Universidad de Buenos Aires, y tiene el fin did\'actico de discutir el significado de las relaciones de incerteza entre tiempo y energ\'{\i}a presentando un brev\'{\i}simo resumen de los trabajos cl\'asicos.

\end{abstract}

\maketitle

\section{Preliminares}

El principio de incerteza entre tiempo y energ\'{\i}a re\-pre\-sen\-ta uno de los ejemplos m\'as concisos de a\-que\-llos aspectos que suelen ser tratados con cierta dis\-pli\-cen\-cia en las discusiones de los libros de texto y cursos introductorios de mec\'anica cu\'antica. Independientemente de las razones de esto, no es dif\'{\i}cil notar que el tratamiento de este tema deja entrever cierto soslayo intencional de los aspectos medianamente sutiles. Esto es as\'{\i} a\'un cuando resulta ser un punto conceptual de cierta importancia en el \'ambito did\'actico precisamente por su propiedad de suscitar confusiones entre quienes se ven frente a \'el por primera vez. Las confusiones se deben principalmente a dos aspectos: { a) La viciosa b\'usqueda de quienes, sin razones demasiado rigurosas, emprenden la piadosa tarea de aunar los papeles que desempe\~nan la variable temporal y los observables correspondientes a la posici\'on en mec\'anica cu\'antica.} { b) Por otro lado, una raz\'on para confusas interpretaciones resulta del poco empleo que del principio de incerteza entre tiempo y energ\'{\i}a se hace en los cursos de mec\'anica cu\'antica, lo cual es fuente de que en muchas ocasiones este tema sea tratado de manera poco profunda}. Esto \'ultimo es, entre poco m\'as, origen de digresiones ociosas al respecto.   

Seg\'un nuestro entender, un tratamiento moderadamente satisfactorio del tema debe, al menos, dar respuesta acabada a las preguntas recurrentes de los alumnos i\-ni\-cia\-dos en la formulaci\'on matem\'atica de la teor\'{\i}a cu\'antica; a saber: {a) ?`Cu\'al es el significado del tiempo $t$ que aparece en el principio de indeterminaci\'on entre tiempo y energ\'{\i}a?} {b) ?`Es factible definir un operador tiempo que permita realizar el observable correspondiente?} { c) Simplemente, ?`incerteza o indeterminaci\'on?}

A pesar de que \'estos no son muchos, es ciertamente posible encontrar algunos textos que tratan el tema con el detalle adecuado. Aqu\'{\i}, a modo de ejemplo, preferimos llamar la atenci\'on sobre la referencia \cite{busch}, que discute variantes de la desigualdad de Heisenberg entre tiempo y energ\'{\i}a de manera precisa y sistem\'atica.

\section{El principio de incerteza de Heisenberg}

\subsection*{Dispersi\'on e incerteza}

En su formulaci\'on original (cf. \cite{heisenberg}), las relaciones de incerteza de Heisenberg aparecen vinculadas a la relaci\'on de conjugaci\'on existente entre ciertos pares de cantidades f\'{\i}sicas; $v.g.$ el tiempo y la energ\'{\i}a. Dicha conjugaci\'on, como bien sabemos, es usualmente entendida como debida al hecho de que ciertos pares de magnitudes intervinientes en la formulaci\'on de la teor\'{\i}a se encuentran relacionadas mediante la transformada de Fourier; esta relaci\'on deviene sin mucho desarrollo en la inecuaci\'on de Heisenberg
\begin{eqnarray}
\Delta E \Delta t \geq \frac {\hbar }{2}  \label{1}
\end{eqnarray}
En parte, debemos esta forma de presentaci\'on a los libros cl\'asicos de texto, los cuales nos explican casos particulares como ejemplos heur\'{\i}sticos para entender tales relaciones. La l\'{\i}nea general de estos argumentos es la siguiente: Si entendemos que un estado f\'{\i}sico de un sistema cu\'antico en un determinado instante $t$ est\'a caracterizado por la funci\'on de onda \footnote{Tomemos por caso un {\it paquete} de ondas como el estado f\'{\i}sico en consideraci\'on} $\psi (t)$ y tenemos presente que el espectro de la distribuci\'on de las componentes de energ\'{\i}a que conforman dicho {\it paquete} de ondas est\'a dado por su transformada de Fourier $\tilde \psi (E/\hbar)$, entonces podemos definir las desviaciones de dichas cantidades seg\'un
\begin{eqnarray*}
(\Delta t)^2&=&\int_{-\infty}^{\infty} dt|\psi (t)|^2(t-t_0)^2 \\
(\Delta E)^2&=&\frac {1}{\hbar } \int_{-\infty}^{\infty} dE|\tilde \psi (E/\hbar )|^2(E-E_0)^2
\end{eqnarray*} 
siendo
\begin{eqnarray*}
t_0=\int_{-\infty}^{\infty} dt|\psi (t)|^2t  \; \; \; \; \;
E_0=\frac {1}{\hbar }\int_{-\infty}^{\infty} dE|\tilde \psi (E/\hbar )|^2E
\end{eqnarray*}
Es un ejercicio {\it standard} de an\'alisis matem\'atico mostrar que de estas definiciones se deduce (\ref{1}) sin mayor dificultad. De esta manera, tenemos la primera y m\'as simple presentaci\'on de las relaciones de incerteza. Seg\'un esto, la relaci\'on (\ref{1}) debe entenderse como {la inecuaci\'on satisfecha por las desviaciones definidas a partir de la distribuci\'on en energ\'{\i}as de los estados constituyentes de un paquete de ondas que depende en el tiempo dada la funci\'on $\psi (t)$}. 

En efecto, esta interpretaci\'on (cf. \cite{bor}) es la adecuada cuando se trata con problemas en los que se refiere a los tiempos caracter\'{\i}sticos de deformaci\'on de un paquete de ondas o cuando se trata del estudio de estados que decaen. Asumiendo conceptos b\'asicos de la mec\'anica cu\'antica, esta es una deducci\'on rigurosa de la inecuaci\'on (\ref{1}) en el caso en el cual se trata con estados del tipo mencionado. No obstante, la intenci\'on de tratar la relaci\'on (\ref{1}) en un contexto m\'as general persiste; discutiremos aqu\'{\i} algunas de las interpretaciones que decoran la bibliograf\'{\i}a.

\subsection*{El tiempo como un operador}

Por otro lado, y lo que lleva en germen quiz\'a la principal fuente de confusi\'on cuando se trata de la energ\'{\i}a y el tiempo, las restantes relaciones de incerteza pueden derivarse en el marco de la formulaci\'on de la teor\'{\i}a en t\'erminos de la teor\'{\i}a  de operadores. De hecho, es usual presentar la deducci\'on de Robertson \cite{r}, quien mostr\'o que la simple consideraci\'on de la desigualdad de Schwarz satisfecha por los vectores de un espacio de estados sobre el cual act\'uan operadores autoadjuntos $A$ y $B$ lleva a una relaci\'on general de la forma 
\begin{eqnarray}
\langle (\Delta A)^2\rangle \langle (\Delta B)^2\rangle  \geq \frac {1}{4} |\langle [A,B]\rangle |^2  \label{robertson}
\end{eqnarray}
y luego, \'esta deviene en
\begin{eqnarray}
\langle (\Delta A)^2\rangle \langle (\Delta B)^2\rangle  \geq \frac {\hbar ^2}{4} \label{usain} 
\end{eqnarray}
si el par de operadores satisfacen $[A,B]=i\hbar $ (cf. \cite{c}). 

Como sabemos, \'esta es la deducci\'on que suele presentarse en los cursos avanzados de mec\'anica cu\'antica; y es, de suyo, el disparador primordial de algunas dudas pertinentes de cualquier alumno medianamente atento que intenta una conciliaci\'on entre la demostraci\'on de (\ref{usain}) y la expresi\'on (\ref{1}). 

El punto central es que ante esta presentaci\'on de las relaciones de incerteza, es ineluctable preguntarse c\'omo entra en este contexto la relaci\'on entre el tiempo y la energ\'{\i}a. En efecto, es dif\'{\i}cil sentirse c\'omodo con el hecho de simplemente asumir que las relaciones de incerteza entre tiempo y energ\'{\i}a son de una ``naturaleza distinta a las restantes'', como podemos leer en alg\'un libro de texto (ver por ejemplo \cite{sakurai}). Aunque, por otro lado, tambi\'en es cierto que una mirada r\'apida a los principios b\'asicos de mec\'anica cu\'antica nos basta para convencernos de que el tiempo entra en escena de una manera distinta a las dem\'as magnitudes. Esto es, si bien es cierto que (\ref{robertson}) nos sugiere casi inmediatamente la idea de iniciar el juego de definir un operador tiempo que represente al observable $t=\left< T\right> $, reconocemos tambi\'en sin demasiada dificultad que, como se\~nal\'o enf\'aticamente Dirac alguna vez, el tiempo en mec\'anica cu\'antica es {\it ab initio} un par\'ametro caracterizado por un n\'umero real $t$ y, por ende, no puede aplicarse la deducci\'on de Roberston a los casos particulares en los cuales esta cantidad se ve involucrada como parte de una relaci\'on de incerteza. 

Es el tiempo un par\'ametro en mec\'anica cu\'antica y no un operador. 
As\'{\i}, la incorporaci\'on de un operador que realice tal observable no puede sino estar caracterizando un observable que refiere a cierto  ``(sub)sistema reloj'' \cite{ab} o, en forma a\'un m\'as gen\'erica, a cierto observable que, dada la naturaleza del problema particular en cuesti\'on, le es dado llevar unidades de tiempo. Lo cierto es que un operador $T$ de tal suerte no puede representar, en t\'erminos generales, ning\'un elemento intr\'{\i}nseco de la descripci\'on mecanocu\'antica. Esto es usualmente mencionado en algunos libros de texto; por ejemplo, en \cite{kimnoz} se enfatiza (como lo hizo originalmente Dirac en \cite{dirac3}) el hecho de que $t$ refiere a un $c$-n\'umero 
en mec\'anica cu\'antica y que, por lo tanto, la cantidad $[t,H]$ es id\'enticamente nula; 
no obstante, la relaci\'on (\ref{1}) es v\'alida a pesar de esto (cf. \cite{dirac}, \cite{dirac2}).

Asimismo, m\'as all\'a de esta aserci\'on, nada nos impide emprender la tarea l\'udica de explorar las implicancias de asumir la existencia de un operador $T$ que satisfaga las siguientes reglas de conmutaci\'on con el hamiltoniano del sistema
\begin{eqnarray}
[H,T]=i\hbar  \label{t}
\end{eqnarray} 
Como primera observaci\'on, notamos que la ecuaci\'on de Heisenberg nos lleva inmediatamente a que
\begin{eqnarray}
\frac {dT}{dt}=\frac {i}{\hbar }[T,H] = 1 \label{cinco}
\end{eqnarray} 
Por lo tanto, (\ref{robertson}) nos permitir\'{\i}a reobtener (\ref{1}) reemplazando $\Delta t$ por el observable correspondiente $\Delta T= \sqrt {\langle (\Delta T)^2\rangle }$ y as\'{\i} reobtendr\'{\i}amos dicha relaci\'on de incerteza como caso par\-ti\-cu\-lar del c\'alculo de operadores. Otro aspecto interesante de la suposici\'on de la existencia de un operador con las propiedades de $T$ es que, en su car\'acter de {\it conjugado} al hamiltoniano, los elementos de matriz $\langle t|\psi _n\rangle $ formados por los vectores $|t\rangle $ del espectro de $T$ (si \'estos existen) y los autoestados $|\psi _n\rangle $ del hamiltoniano tendr\'{\i}an la forma $\langle t|\psi _n\rangle  \sim e^{i\frac {E_n t}{\hbar }}$, que es precisamente la dependencia temporal de la funci\'on de onda, de manera an\'aloga a como los elementos $\langle x|\psi _n\rangle $ representan la dependencia en t\'erminos de la posici\'on.

Sin embargo, m\'as all\'a de las digresiones relacionadas a la definici\'on del operador $T$, Pauli elev\'o al rango de teorema la observaci\'on de que la existencia de un operador que satisfaga (\ref{t}) y que adem\'as posea un espectro continuo implicar\'{\i}a que el espectro del hamiltoniano no sea discreto y acotado inferiormente \cite{paula}. Esto es b\'asicamente debido a que $T$ generar\'{\i}a en ese caso las traslaciones en el espectro de $H$, las cuales estar\'{\i}an caracterizadas por un par\'ametro real que recorrer\'{\i}a dicho espectro. De esta manera, vemos que no es posible incluir en la descripci\'on, para el caso general, un operador herm\'{\i}tico $T$ que represente al tiempo de manera tan ingenua. No obstante, perm\'{\i}tasenos mencionar el cl\'asico ejemplo particular del operador \footnote{Obs\'ervese que otros ordenamientos de este operador sa\-tis\-fa\-cen las mismas propiedades}
\begin{eqnarray}
T=\frac {m}{2} \{X,P^{-1}\} \label{este}
\end{eqnarray}  
el cual, adem\'as de ser un producto sim\'etrico de potencias de operadores localmente autoadjuntos, resuelve la ecuaci\'on $[H,T]=i\hbar $ para el hamiltoniano de una part\'{\i}cula libre de masa $m$. Volveremos a mencionar este operador en las pr\'oximas secciones cuando mencionemos la cr\'{\i}tica de Aharonov y Bohm a las interpretaciones cl\'asicas de la relaci\'on (\ref{1}). Es importante notar la singularidad existente en el espectro de $P^{-1}$ en (\ref{este}). Otros ejemplos que permiten soslayar la prohibici\'on impl\'{\i}cita por el teorema de Pauli son aqu\'ellos que corresponden a hamiltonianos que no est\'an acotados inferiormente. Tal es el caso del hamiltoniano de una part\'{\i}cula gravitante, sometida al potencial $V(x)=m g \ X$, que admite $T= \frac {1}{mg} P$ como definici\'on del operador que realiza (\ref{cinco}), siendo $mg$ una constante del problema. La discusi\'on del ``tiempo como un operador'' no puede considerarse completa sin referir al art\'{\i}culo \cite{Prigogine}, el cual se discute el tema de una interesante forma; en particular, se comenta la relaci\'on existente entre la definici\'on de un ``operador tiempo'' y la definici\'on de un observable correspondiente a la entro\'{\i}a.

\section{Interpretaciones lockianas del principio de incerteza}

\subsection*{La deducci\'on de Mandelstam-Tamm}

Volviendo a la interpretaci\'on de la relaci\'on (\ref{1}) en el contexto de la descripci\'on de estados tipo ``paquetes'' que decaen o se deforman en el tiempo, cabe mencionar con particular atenci\'on la celebrada deducci\'on que Mandelstam y Tamm presentan en \cite{mt}. En ese art\'{\i}culo, los autores comienzan se\~nalando la existencia de una ``conexi\'on general entre la dispersi\'on del espectro de energ\'{\i}as de un cierto estado y la permanencia en el tiempo de sus magnitudes f\'{\i}sicas'', caracterizadas \'estas por los observables del sistema. Ellos se valen de dicha ``conexi\'on'' para definir una formulaci\'on cuantitativa de la relaci\'on (\ref{1}). El an\'alisis de \cite{mt} comienza con la consideraci\'on de la ecuaci\'on de Ehrenfest-Heisenberg para un dado operador $A$
\begin{eqnarray}
\langle \frac {dA}{dt}\rangle =\frac {i}{\hbar }\langle [A,H]\rangle 
\end{eqnarray}
Luego, teniendo en cuenta (\ref{robertson}), se llega a mostrar que se satisface la siguiente relaci\'on entre cocientes incrementales
\begin{eqnarray}
\langle \Delta H\rangle \delta t  \geq \frac {\hbar }{2} \frac {\langle A(t+\delta t)-A(t)\rangle }{\delta A}   
\end{eqnarray}
donde $\delta A$ refiere al valor promedio que la cantidad $\Delta A $ adquiere en el intervalo de tiempo infinitesimal $\delta t$. Se deduce entonces la siguiente f\'ormula
\begin{eqnarray}
\langle \Delta H\rangle \Delta t  \geq \frac {\hbar }{2}  
\end{eqnarray}
donde $\Delta t$ es el valor que minimiza el intervalo de tiempo $\delta t $ en el cual el valor medio de cierta cantidad $A$ se ve modificado en una cantidad igual a su promedio (ver \cite{mt} para los detalles).

Un tratamiento an\'alogo al presentado por Mandelstam y Tamm es descripto en la referencia \cite{s2} por Shalitin. En este caso, se define la cantidad siguiente
\begin{eqnarray}
 \Delta T  = \frac {\Delta A}{\frac {d}{dt}|\langle A\rangle |}  
\end{eqnarray}
donde, por supuesto, $\Delta A$ y $\langle A\rangle $ est\'an referidos a un estado particular del sistema. As\'{\i} definido, $\Delta T$ {mide el intervalo de tiempo en el cual la cantidad} $\left< A\right> $ {es confinada en un intervalo de incerteza $\Delta A$ en torno a su valor medio}. Esto es, el an\'alisis presentado en \cite{s2} (al igual que el presentado en \cite{mt}) se basa en la definici\'on de una medida de la ``identidad'' del estado que evoluciona en el tiempo \footnote{En este sentido, podemos referirnos al car\'acter lockiano de las definiciones de los art\'{\i}culos \cite{mt} y \cite{s2}}.

Shalitin trata a modo de ejemplo una aplicaci\'on de esta forma de interpretar la relaci\'on de incerteza entre tiempo y energ\'{\i}a al caso de estados metaestables caracterizados por el operador proyector $A=|\psi \rangle \langle \psi |$. En este an\'alisis, $|\psi \rangle $ es un estado del sistema que satisface, en alguna aproximaci\'on, la relaci\'on $|\frac {d}{dt} \langle A\rangle |=\frac {1}{\tau }\langle A\rangle $ dado que se tratan en consideraci\'on estados con un comportamiento de la forma $\langle A\rangle \sim e^{-t/\tau }$. Luego, el simple reemplazo del comportamiento del estado metaestable en la definici\'on de $\Delta T$ de arriba lleva a obtener que en el l\'{\i}mite de largos tiempos se cumple 
\begin{equation}
\Delta H \tau \sim \hbar . \label{simi}
\end{equation}
Shalitin comenta luego la comparaci\'on con la interpretaci\'on usual basada en la regla de oro de Fermi para transiciones \cite{landau3}. As\'{\i}, {aparece el tiempo caracter\'{\i}stico $\tau $ de decaimiento del estado metaestable como cantidad interviniente en una relaci\'on de equivalencia que involucra a la incerteza $\Delta H$}.

La relacion (\ref{simi}) es usualmente referida en las aplicaciones en f\'{\i}sica de part\'{\i}culas, donde los tiempos de vida media $\tau $ de estados inestables aparecen en relaci\'on con el principio de Heisenberg. No obstante, notemos que elementos adicionales, tales como la forma particular de la evoluci\'on de los estados, debe ser asumida para reducir la inecuaci\'on de Heisenberg a una relaci\'on de identidad (realizada por el s\'{\i}mbolo $\sim $) que involucre el tiempo medio $\tau $. Es a esto a lo que nos habituamos cuando estudiamos la relaci\'on (\ref{1}) como relacionada con la regla de oro de Fermi en el desarrollo de teor\'{\i}a de perturbaciones.

\subsection*{Sobre otras interpretaciones}

Por \'ultimo, nos permitimos llamar la atenci\'on sobre una interesante derivaci\'on de la desigualdad (\ref{1}) empleando el formalismo de matriz densidad. Esta deducci\'on se debe originalmente a Eberly y Singh \cite{es} y fue did\'acticamente desarrollada en \cite{b} por Blanchard. En este contexto, la incerteza en el tiempo $\Delta t$ est\'a medida en t\'erminos de la derivada temporal de la matriz densidad en el {\it picture} de Schr\"odinger, siendo 
\begin{equation}
(\Delta { \tilde {t}} )^{-2} = \frac 14 \left< \left( \frac {\partial \rho }{\partial t} \right)^2 \right>
\end{equation}
que satisface la desigualdad (\ref{1}) para la incerteza en la energ\'{\i}a $\Delta E$, definida \'esta seg\'un
\begin{equation}
(\Delta E)^2 = \left< \left( H-\left< H\right> \right) ^2 \right> ,
\end{equation}
y es posible mostrar que una desigualdad del tipo (\ref{1}) se cumple entre $\Delta \tilde t$ y $\Delta E$. Esta descripci\'on propone un criterio de selecci\'on ya que seg\'un este an\'alisis los estados puros, a diferencia de los estados mixtos, satisfacen la m\'{\i}nima relaci\'on de incerteza, {\it i.e.} $\Delta E \Delta \tilde t = \hbar /2$. Asimismo, esto nos advierte de la diferencia que existe con la desigualdad de Heisenberg propiamente dicha, ya que no es dif\'{\i}cil idear ejemplos en los que la desigualdad estricta $\Delta E \Delta t > \hbar /2$ se verifica a\'un para estados puros. Volvamos, en este punto, a llamar la atenci\'on sobre la referencia \cite{Prigogine}.

\section{Los procesos de medici\'on y relaci\'on de incerteza}

\section*{La cr\'{\i}tica de Aharonov-Bohm}

En un reconocido art\'{\i}culo de Landau y Peierls \cite{lp} (ver tambi\'en \cite{landau4}), en el cual se ensayaba tempranamente sobre las limitaciones que est\'an impl\'{\i}citas en el intento por extender al rango relativista las cantidades f\'{\i}sicas definidas en la mec\'anica ondulatoria, se concluye la existencia de ciertas limitaciones \footnote{Los autores discuten la imposibilidad de satisfacer la llamada {\it condici\'on de repetitividad} en el contexto de la mec\'anica cu\'antica relativista} deducidas del hecho de asumir que la energ\'{\i}a no puede ser medida con arbitraria exactitud en un corto lapso de tiempo debido a la afecci\'on provocada por el mismo proceso de medici\'on a\'un en el caso de mediciones predecibles. Los autores afirman esto escud\'andose en la referencia expl\'{\i}cita al punto de vista de Bohr al respecto. 

As\'{\i}, la interpretaci\'on sugerida en \cite{lp} descansa en la idea de que en un tiempo $\Delta t$ no puede hacerse una medici\'on en la energ\'{\i}a de un sistema para la cual la discriminaci\'on sea menor \footnote{Landau y Peierls concluyen expl\'{\i}citamente que no puede existir una medici\'on predecible en la mec\'anica ondulatoria salvo si se trata de cantidades constantes en el tiempo. Cabe mencionar que en la referencia \cite{landau4} se detallan los re\-sul\-ta\-dos \cite{lp} enfatizando la relaci\'on entre el principio de incerteza y la existencia de la velocidad de la luz en su car\'acter de velocidad m\'axima} que $\hbar / \Delta t$.

Expuesto este punto de vista, seg\'un el cual se propone la interpretaci\'on de la inecuaci\'on (\ref{1}) es que {existe un l\'{\i}mite en la determinaci\'on de la medici\'on de la energ\'{\i}a relacionada con el tiempo de duraci\'on de dicha medici\'on}, debemos mencionar el embate cr\'{\i}tico que Aharonov y Bohm iniciaron al respecto en \cite{ab}. En este art\'{\i}culo, se critica la interpretaci\'on de Landau y Peierls arguyendo que una afirmaci\'on semejante no puede estar impl\'{\i}cita en la formulaci\'on matem\'atica de la teor\'{\i}a cu\'antica y que, por lo tanto, no puede ser considerada como la interpretaci\'on adecuada de (\ref{1}). Tambi\'en se\~nalan que los ejemplos presentados en la literatura ($e.g.$ en las referencias \cite{lp} y \cite{mt}) no son lo suficientemente generales como para inferir a partir de ellos la interpretaci\'on final.
A modo de ep\'{\i}logo de su trabajo, Aharonov y Bohm presentan como contraejemplo un proceso de medici\'on en el cual puede medirse la energ\'{\i}a de un sistema en un tiempo finito y con arbitraria exactitud. 

En \cite{ab} se indica tambi\'en que la interpretaci\'on err\'onea de (\ref{1}) se debe a una mala lectura del punto de vista de Bohr y se enfatiza que, de ser cierta una relaci\'on de incerteza entre el tiempo de duraci\'on de una medici\'on y la precisi\'on de la misma, \'esta deber\'{\i}a ser demostrable a partir del formalismo de la teor\'{\i}a ($v.g.$ en t\'erminos del c\'alculo de operadores). As\'{\i}, ellos tratan el caso particular del operador $T$ definido en (\ref{t}) como aqu\'el que corresponde al observable del tiempo medido por el aparato de medici\'on, acentuando la diferencia entre el tiempo del aparato de medici\'on y el tiempo {\it interno} del sistema \footnote{Hasta donde alcanzamos a ver, la discusi\'on original de Aharonov y Bohm no cierra satisfactoriamente este punto dada una pobre definici\'on formal de lo que ellos denominan tiempo $interno$}; este \'ultimo conmuta con el hamiltoniano del aparato de medici\'on. Seg\'un este an\'alisis, (\ref{1}): la relaci\'on es satisfecha para los observables del reloj.

{La cr\'{\i}tica de Aharonov y Bohm \footnote{Hoy podemos encontrar en la literatura renovadas discusiones acerca de la interpretaci\'on del principio de incerteza entre tiempo y energ\'{\i}a para sistemas cerrados haciendo nuevamente hincapi\'e en la incerteza del tiempo como debida a la medici\'on de una cierta porci\'on del sistema considerada como $reloj$ en un contexto an\'alogo al presentado en \cite{ab}.} deja constancia de que la interpretaci\'on de la relaci\'on (\ref{1}) como relacionada al tiempo y a la precisi\'on de una determinaci\'on en el valor de la energ\'{\i}a es, si no incorrecta, s\'olo v\'alida en casos particulares celosamente elegidos.} Y es esta nuestra primera conclusi\'on. La que, por su parte, hace a la cuesti\'on terminol\'ogica entre incerteza e indeterminaci\'on. A saber: {el principio de Heisenberg no est\'a vinculado con la indeterminaci\'on en el proceso de medici\'on, sino con la incerteza cu\'antica intr\'{\i}nseca de la funci\'on de onda y el espectro de energ\'{\i}a asociado a dicha funci\'on.}

\subsection*{Sobre el proceso de medici\'on de un observador interno}

Pocos a\~nos atr\'as, Aharonov y Reznik retomaron en \cite{ab2} la pregunta acerca de si hay alguna conexi\'on entre la desigualdad (\ref{1}) y el acto de la medici\'on de la energ\'{\i}a. En el contexto del art\'{\i}culo \cite{ab2}, a diferencia de los casos tratados en \cite{ab}, el proceso de medici\'on de la energ\'{\i}a y el tiempo empleado en tal proceso refieren a una medici\'on que se realiza desde el mismo sistema; esto es, una medici\'on realizada por un ``observador interno''. Aharonov y Reznik arguyen que en los ``sistemas cerrados'' es posible asociar una relaci\'on del tipo (\ref{1}) al proceso de (auto)medici\'on de la energ\'{\i}a $E$, dada con una precisi\'on $\Delta E$, si se realiza tal estimaci\'on en un intervalo $\Delta t$ (cf. \cite{e}, \cite{ab3}). En palabras de los autores, se lee ``{\it This time-energy uncertainty 
principle for a closed system follows from the measurement back-reaction on 
the system.}''.

\section{Otras digresiones al respecto}

\subsection*{La discusi\'on en el contexto de las teor\'{\i}as relativistas}

Siguen en la lista de las discusiones m\'as frecuentes en la literatura referidas a la interpretaci\'on de (\ref{1}) aqu\'ellas que se basan en esa piadosa b\'usqueda de razones que permitan al tiempo y a las coordenadas esenciarse en la formulaci\'on de la mec\'anica cu\'antica. Mencion\'abamos esto anteriormente. Estas discusiones se centran muy frecuentemente en el mito de que la asimetr\'{\i}a que establece el papel particular del tiempo en la formulaci\'on original implica de alguna ma\-ne\-ra una incompatibilidad con los principios de covariancia de la relatividad especial. Por supuesto que esta idea es falaz; bien sabemos que {la entera formulaci\'on de la mec\'anica cu\'antica entra en el marco de las teor\'{\i}as relativistas de campos sin que sea el papel privativo del tiempo un riesgo para la invariancia de Lorentz.} 

No obstante, no debe entenderse de esto que no se reconoce que la exploraci\'on del significado de los aspectos cu\'anticos en el \'ambito de la relatividad especial sea un tema de merecida atenci\'on. Existen diversos e interesantes art\'{\i}culos que tratan tales temas. Por ejemplo, Hilgevoord se\~nala en \cite{h} aspectos sutiles de la interpretaci\'on de la variable temporal $t$ en comparaci\'on a una confusi\'on usual que \'el remarca entre las coordenadas espaciales y los observables de posici\'on (cf. \cite{hkn}). B\'asicamente, lo que Hilgevoord propone es que la simetr\'{\i}a entre la variable temporal y las coordenadas espaciales est\'a, de hecho, manifiesta debido precisamente a que en ambos casos la desigualdad de Heisenberg refiere a los par\'ametros $t$ y $x$ y no al hecho de que alguno de ellos est\'e eventualmente relacionado con un operador que lo realice. Esto se vincula con la idea de que, al intentar formular una teor\'{\i}a relativista de part\'{\i}culas, se llega a la teor\'{\i}a de campos donde, sin abandonar ning\'un precepto de la formulaci\'on cu\'antica, surge ``el campo'' $\Phi (x,t)$ como protagonista principal, el cual es ahora ``el ente a ser cuantizado''. As\'{\i}, las coordenadas, relegadas al papel de par\'ametros de la variedad donde el campo se formula, acompa\~nan a la definici\'on del nuevo par conjugado $\Pi (x,t), \Phi (x,t)$.

Siguiendo con nuestro recorrido bibliogr\'afico, pueden encontrarse en la literatura diversos ensayos sobre extensiones del principio de incerteza al caso relativista proponiendo relaciones entre el tiempo propio y el observable correspondiente a las masas de las part\'{\i}culas, de la forma $\Delta \tau \Delta m \sim \hbar /2c^2$.

Tambi\'en es frecuente verse frente a otros argumentos basados en la consideraci\'on del campo gravitatorio ya que, de hecho, algunos aspectos privativos de las teor\'{\i}as de gravedad ampl\'{\i}an en gran medida la gama de aspectos interesantes relacionados con las relaciones de incerteza: La inclusi\'on del tiempo de Planck $t_{Planck} = \sqrt {\frac {G \hbar}{c^5}}$ como nueva escala en la teor\'{\i}a, el advenimiento de las teor\'{\i}as formuladas sobre geometr\'{\i}as no conmutativas y la consideraci\'on de objetos fundamentales extendidos como cuerdas o D-branas son ejemplos de nuevos elementos que proponen un feraz y nuevo terreno para el estudio de estos temas en la f\'{\i}sica te\'orica.

Puebla la literatura la gama m\'as diversa de discusiones relacionadas, que van desde interesantes tratados de interpretaci\'on hasta las m\'as extravagantes digresiones.

\subsection*{La falacia de la incerteza cu\'antica como margen para la conservaci\'on de la energ\'{\i}a}

En otro contexto, resulta interesante comentar algunas otras disquisiciones tambi\'en relacionadas con el principio de Heisenberg. Es \'este el caso de un argumento heur\'{\i}stico que, a\'un cuando falto de rigor, es usualmente utilizado para apuntalar la interpretaci\'on de los procesos de interacci\'on entre part\'{\i}culas fundamentales como promovida por el intercambio de part\'{\i}culas virtuales que ``viven'' un tiempo menor al intervalo $\Delta t \sim \frac {\hbar}{\Delta E}$ a costa de una nimia violaci\'on de la energ\'{\i}a en la cantidad $\Delta E$ exigida por su propia existencia. Esta imagen bosquejada de lo que luego se formaliza en t\'erminos de la electrodin\'amica cu\'antica, establece que las part\'{\i}culas virtuales ({\it v.g.} fotones virtuales) que portan la interacci\'on ({\it resp.} electromagn\'etica) entre las part\'{\i}culas cargadas deben su e\-xis\-ten\-cia a una violaci\'on de la conservaci\'on de la energ\'{\i}a que s\'olo ocurre en un per\'{\i}odo de tiempo protegido por el principio de incerteza. Si bien no es dif\'{\i}cil mostrar que esta imagen de los hechos es incorrecta (o al menos incontrastable), vale en muchos casos como ilustraci\'on, y hasta resulta eficaz para obtener informaci\'on de las interacciones a partir de ella. Por ejemplo, si consideramos que la interacciones entre dos electrones est\'a mediada por la emisi\'on de uno de estos fotones virtuales que transmite una energ\'{\i}a de interacci\'on $\Delta E$, y suponemos que la vida de este portador es $\Delta t$ del orden de
\begin{eqnarray*}
\Delta E \Delta t \sim {\hbar } ,
\end{eqnarray*}
entonces nos basta tener en cuenta que el fot\'on viaja a la velocidad de la luz $c$ y que en ese tiempo puede recorrer una distancia $r=c\Delta t$ para obtener que la energ\'{\i}a total intermediada en el proceso estar\'{\i}a dada por $E(r)=\frac {e^2}{\hbar c} \Delta E$, ya que $\frac {e^2}{\hbar c} $ refiere a la probabilidad de que este tipo de proceso ocurra \footnote{Como muestra el c\'alculo a primer orden en la electrodin\'amica cu\'antica.}, siendo $e$ la carga el\'ectrica del electr\'on. De esta manera, la energ\'{\i}a mediada en la interacci\'on estar\'a dada por
\begin{equation}
E \sim \frac {e^2}{r}   \label{coulomb}
\end{equation}
que precisamente coincide con el potencial el\'ectrico (cf. \cite{ha}). Siguiendo las referencias de la literatura es posible toparse con art\'{\i}culos en los cuales se aventuran extrapolaciones de estas ideas a terrenos tales como el de la cosmolog\'{\i}a \cite{tyron} o la f\'{\i}sica hadr\'onica \cite{ha}.

Hay una inevitable conclusi\'on a la que se arriba luego de explorar las diversas aristas del problema de la interpretaci\'on del principio de incerteza entre tiempo y energ\'{\i}a. Esta conclusi\'on hace a la innumerable cadena de falacias en la que se ve envuelto dicho principio cuando se lo trata en un contexto did\'actico. Dichas falsas interpretaciones son promovidas, de manera exclusiva, por un trato imprudente del tema. Un ejemplo conciso del alcance de dichas digresiones es la referida arriba acerca de la extrapolaci\'on exagerada de la incerteza cu\'antica como motivo para la existencia de los ``portadores virtuales'' en los procesos de interacci\'on. De hecho, no hace falta un doctorado en l\'ogica formal para reconocer que ninguna ley de la f\'{\i}sica basada en una ecuaci\'on ({\it e.g.} la ley de Coulomb) puede ser derivada de una inecuaci\'on como (\ref{1}) sin agregar, aunque m\'as no sea entre l\'{\i}neas, hip\'otesis adicionales. Sobre este punto, valga otra objeci\'on: Si verdaderamente el potencial (\ref{coulomb}) puede ser obtenido a partir de la consideraci\'on del principio de Heisenberg y poco m\'as, entonces surge la pregunta acerca de c\'omo se explica que no haya una derivaci\'on an\'aloga del potencial coulombiano en un espacio de dimensionalidad gen\'erica $d$ (que bien sabemos que est\'a dado por $E \sim \frac {e^2}{r^{d-2}}$, \footnote{Notemos que los argumentos cinem\'aticos empleados no son dependientes de la dimensionalidad del espacio plano que se considere.}). Si estos argumentos sobrevivieran deber\'{\i}amos, pues, inferir que el principio de Heisenberg nos habla de la dimensionalidad del espacio y de las cualidades del caso $d=3$, y claro que todo esto no tiene sentido alguno.

Las confusiones conceptuales que est\'an en germen en este tipo de argumentos se ponen de manifiesto con razones similares a las que usamos para explicar el hecho de que coexistan tantas y tan diversas interpretaciones de (\ref{1}). A saber: {El principio de incerteza de Heisenberg, entendido \'este como un car\'acter intr\'{\i}nseco de la formulaci\'on de la teor\'{\i}a cu\'antica y, por ende, independiente del problema tratado en cada caso particular, no puede sino referir a elementos b\'asicos de la teor\'{\i}a tales como la existencia de una magnitud fundamental $\hbar $, la ecuaci\'on de Schr\"odinger o las propiedades algebraicas del hamiltoniano y del espacio de estados sobre el cual \'este act\'ua}. Es por eso que la inclusi\'on de artefactos tales como observables con dimensiones de tiempo representados por operadores autoadjuntos as\'{\i} como magnitudes de escalas t\'{\i}picas de dispositivos de medici\'on no pueden llevar {\it per se} la raz\'on de ser ni el significado de (\ref{1}). Esto es, el significado de (\ref{1}) no puede depender de la carga del electr\'on, ni de la velocidad de la luz, ni de la escala de Planck.

\subsection*{A modo de corolario: Magnitudes y escalas}

En efecto, si hay alg\'un vestigio de verdad en los sofismas atacados anteriormente, \'este no puede estar relacionado m\'as que con la casualidad o con el simple hecho de que s\'olo una verdad de perogrullo se esconde atr\'as de atribuirle a (\ref{1}) un nuevo significado. 
El significado del principio de incerteza de Heisenberg es aquel que se relaciona con las campanas de dipersi\'on de las distribuciones en el espectro de energ\'{\i}as que discutimos al comienzo. Su significado est\'a claro en los casos en los cuales se trata con configuraciones que nos permiten hablar de ``paquetes de onda'' y es lo que se discute, por ejemplo, en las interpretaciones {\it alla} Mandelstam y Tamm.

Por supuesto que cuando la pregunta acerca del significado de cierta expresi\'on se dirige hacia una teor\'{\i}a fundamental y tan de base como la mec\'anica cu\'antica, la respuesta habr\'{\i}a de resultar m\'as f\'acil. Esto es as\'{\i} por cu\'antos menos elementos hay de los cuales puede depender el significado de lo que se desea entender (en este caso, el significado de (\ref{1})). Por esto, cualquier significado del principio de incerteza debe estar, como se\~nalaron tempranamente Aharonov y Bohm, en la descripci\'on matem\'atica de la teor\'{\i}a. 

Por su parte, dada la austeridad a la hora de contar los elementos b\'asicos en la formulaci\'on de la teor\'{\i}a cu\'antica \footnote {y m\'as espec\'{\i}ficamente, dada la escasez de cantidades fundamentales que involucren escalas temporales y de energ\'{\i}a a parte de $\hbar$}, no debe resultar demasiado sorprendente que la sola inclusi\'on de unas pocas cantidades con dimensiones de tiempo y energ\'{\i}a a la hora de atacar un problema particular nos lleve a obtener relaciones del tipo $\Delta t \Delta E \sim \hbar$, o bien $\Delta t \Delta E < \hbar$, o bien $\Delta t \Delta E > \hbar$, o bien $\Delta t \Delta E = 177 \hbar$; pero esto no es m\'as que una casualidad cuya frecuencia es bien explicada por la generalidad y frugalidad del formalismo, lo que lleva a que no haya, {\it ab initio}, demasiadas magnitudes con unidades de {\it tiempo} $\times $ {\it energ\'{\i}a}.

En resumen, no debemos interpretar que cada relaci\'on que lleve tiempos y energ\'{\i}as del lado izquierdo y una constante relacionada con $\hbar$ del lado derecho resulta ser una manifestaci\'on hasta entonces desconocida del principio (\ref{1}). Basta para convencernos de esto considerar otro ejemplo: Recientemente se ha arg\"u\'{\i}do que la cuantizaci\'on can\'onica de la gravedad lleva a que los estados puros evolucionan naturalmente hasta convertirse en estados mixtos debido a una decoherencia inducida por la no-existencia de relojes ideales cl\'asicos, los cuales son reemplazados en esta teor\'{\i}a por ``relojes cu\'anticos'' \cite{pullin}. As\'{\i}, aparecer\'{\i}a en este tratamiento de la gravedad cu\'antica una escala de tiempos de decoherencia $t_{decoh}$ que, combinada con la escala de Planck $t_{Planck}$ para definir la cantidad $(\Delta t )^2 = t_{decoh} t_{Planck}$, deviene en una relaci\'on
\begin{equation*}
\Delta t \Delta E \sim \hbar
\end{equation*} 
donde $E / \hbar $ es la frecuencia asociada a la dispersi\'on en el espectro de energ\'{\i}as de los estados del sistema que est\'a bajo estudio. Y esto tiene poco (si no es que absolutamente nada) que ver con (\ref{1}). En efecto, esto resulta en una modificaci\'on de la mec\'anica cu\'antica en uno de sus basamentos: el car\'acter del tiempo. Por lo cual una conexi\'on de esto con el principio de Heisenberg es, si no nula, para nada evidente.

\section{Sobre la naturaleza del tiempo en mec\'anica cu\'antica}

\subsection*{El tiempo cosmol\'ogico}

Retomemos ahora el tema de los observables de tiempo. La inclusi\'on de operadores temporales $T$ que reemplacen al $c$-n\'umero $t$ con el que la mec\'anica cu\'antica naci\'o no es adecuada m\'as all\'a de los (no tan generales) ejemplos en los cuales se le puede asignar el papel de reloj a alguna parte del sistema que rige cierto per\'{\i}odo de alg\'un subproceso ({\it e.g.} de medici\'on). 


Para precisar esto, demos un ejemplo de contexto en el cual la b\'usqueda de un operador $T$ que realice el observable de tiempo en la teor\'{\i}a cu\'antica adquiere sentido: \'este es el caso del programa de cuantizaci\'on de modelos cosmol\'ogicos provenientes de la acci\'on de Einstein-Hilbert para el campo gravitatorio. La gravedad es un ejemplo de lo que se conoce como modelo hamiltoniano con un v\'{\i}nculo cuadr\'atico y, en particular, el v\'{\i}nculo existente en este caso se traduce en el hecho de que el hamiltoniano $H$ se anula id\'enticamente, {\it i.e.} $H = 0$.

La versi\'on cu\'antica de tal ecuaci\'on se conoce con el nombre de ecuaci\'on de Wheeler-De Witt. Luego, la cuantizaci\'on de tal tipo de teor\'{\i}a requiere como paso previo la identificaci\'on de un operador temporal $T$ \footnote{que estar\'a dado en t\'erminos de los grados de libertad del campo gravitatorio $g_{ij}$ y de sus variables can\'onicas conjugadas $\pi _{ij}$.} que permita ser identificado con {\it el tiempo del sistema} y que satisfaga estar globalmente bien definido. Esto \'ultimo es, entre otras cosas, pedir que el operador $[H,T]$ sea definido positivo sobre el espacio de funciones de onda $\psi$, que adquieren en este contexto la interpretaci\'on de {\it funciones de onda del universo} \cite{cosmo}. En muchos de los modelos cosmol\'ogicos cu\'anticos que representan universos homog\'eneos, el operador tiempo $T$, que siempre est\'a vinculado a los grados de libertad de la geometr\'{\i}a (universo) en cuesti\'on, es directamente identificado con el radio del universo en expansi\'on u otra variable asociada, {\it e.g.} alguna medida de la anisotrop\'{\i}a, {\it etc}. As\'{\i}, cuando la pregunta se refiere al universo en su totalidad el par\'ametro temporal no es tratado como un elemento externo (ver \cite{Simeone} para un {\it review} de este tema).

Ahora bien, peculiaridades tales como la representaci\'on del tiempo en la cuantizaci\'on de modelos cosmol\'ogicos resultan ser un d\'ebil argumento como para extrapolar semejante realizaci\'on al terreno de la teor\'{\i}a cu\'antica en un contexto general. En todo caso, cualquier intento por hacerlo conforma, en s\'{\i}, una generalizaci\'on de la teor\'{\i}a. 

\subsection*{El tiempo y los relojes}

Siguiendo con esta observaci\'on, hay en la literatura intentos por escindir al tiempo de su car\'acter de {par\'ametro independiente de los procesos f\'{\i}sicos que transcurren en \'el}. Se ensay\'o la posibilidad de {\it rever} el concepto de evoluci\'on temporal en mec\'anica cu\'antica bas\'andose en la referencia a un subproceso f\'{\i}sico que oficia, de este modo, de {\it elemento reloj}. Por ejemplo, Wootters trata en \cite{wootters} el ejemplo de un sistema de part\'{\i}culas, el cual le sirve para mostrar que el la evoluci\'on temporal descrita en t\'erminos del par\'ametro $t$ es reemplazable por la correlaci\'on cu\'antica entre las distintas part\'{\i}culas del sistema usando una de ellas como ``reloj''. Con grandilocuencia, Wootters afirma como conclusi\'on que ``no es necesario incluir al tiempo como un elemento b\'asico en la descripci\'on del mundo''.

\section{Conclusiones}

Condensando, pues, en forma de corolarios aquellas conclusiones que derivamos, podemos enunciar los siguientes:

{\it a)} {La controversia terminol\'ogica entre incerteza e indeterminaci\'on se plantea, de forma m\'as precisa, en t\'erminos de la pregunta acerca de si es correcta la interpretaci\'on del principio de Heisenberg (\ref{1}) como involucrando a la precisi\'on en una medici\'on de la energ\'{\i}a realizada en un intervalo $\Delta t$ y que arroja un resultado con indeterminaci\'on $\Delta E$. La respuesta a esta pregunta es negativa y ha sido expuesta en el trabajo de Aharonov y Bohm con lucidez}. Es esta una de las m\'as frecuentes confusiones.

{\it b)} {Acerca de la cuesti\'on de si existe un operador tiempo que realice el observable correspondiente de manera de aunar (\ref{1}) con los casos que son tratables con la deducci\'on de Robertson, podemos responder lo siguiente: La existencia de un operador temporal $T$ en mec\'anica cu\'antica no es una propiedad general de todo caso estudiado. Cierto es que hay ejemplos en los cuales es factible definir un operador de tal suerte; no obstante, en tales casos, y como resulta evidente, el observable $\left< T \right>$ refiere a una cantidad particular y propia de dicho ejemplo y no es el tiempo $t$ que la mec\'anica cu\'antica contempla en sus fundamentos.} La existencia de $T$ tal que $\left< T \right>$ resulte mon\'otono en el tiempo $t$ es una particularidad del hamiltoniano particular bajo estudio y no representa una raz\'on para pretender que $t$ resulte reemplazado en la formulaci\'on de la teor\'{\i}a cu\'antica. Las sutilezas del ejemplo de la cosmolog\'{\i}a cu\'antica son desarrolladas con pericia en la literatura.

{\it c)} {Tambi\'en podemos dar una raz\'on para la frecuente aparici\'on de falaces interpretaciones de la desigualdad de Heisenberg. Este fen\'omeno se debe, como mencion\'abamos, a la simplicidad de los ejemplos tratados usualmente que, sumada a la austeridad del formalismo de la teor\'{\i}a, hacen f\'acil incurrir en el error de relacionar cualquier relaci\'on del tipo $\Delta t \Delta E \sim \hbar$ con la desigualdad (\ref{1}). Por supuesto, cuando los ejemplos y modelos estudiados aumentan en complejidad y adquieren m\'as estructura (}{\it i.e.} {m\'as elementos con unidades en un sistema) las relaciones funcionales involucrando tiempos y energ\'{\i}as caracter\'{\i}sticas se multiplican}. Para expresar esto en forma a\'un m\'as concisa: La discusi\'on del principio de incerteza mediante ejemplos particulares introduce escalas de energ\'{\i}a y de tiempos que son propios de dichos ejemplos. As\'{\i}, algunas relaciones entre dichos tiempos y energ\'{\i}as pueden satisfacer relaciones an\'alogas a (\ref{1}). No obstante, extrapolar conclusiones a partir de esto es riesgoso ya que puede llevar a errores de interpretaci\'on. El mejor ejemplo es, acaso, el de los m\'etodos ideados para medir la energ\'{\i}a de una part\'{\i}cula libre bas\'andose en el ``tiempo de vuelo'' de \'esta, lo que llev\'o al error de asociar (\ref{1}) al acto de medici\'on; interpretaci\'on que, como mencionamos, fue criticada en \cite{ab} con agudeza. 

{\it d)} Por \'ultimo, enfaticemos que no hay forma de deducir identidades del tipo $\Delta E \tau \sim \hbar$ a partir de (\ref{1}) sin asumir elementos adicionales. Esto es tan cierto cuanto que no hay conexi\'on directa entre el principio de Heisenberg y una licencia para la violaci\'on de las leyes de conservaci\'on en la naturaleza.

De esta manera, reconocemos que a\'un en aquellos momentos en los cuales nos permitimos disfrutar de los argumentos heur\'{\i}sticos y tratamientos cualitativos de los fen\'omenos, no debemos perder de vista que toda consideraci\'on seria en f\'{\i}sica debe estar sustentada, casi por definici\'on, por una formulaci\'on precisa de los entes intervinientes en t\'erminos de relaciones matem\'aticas y, sumado a esto, de una espec\'{\i}fica estructura sem\'antica que d\'e cuenta del significado un\'{\i}voco de cada representaci\'on.

\end{document}